\documentclass[%
 reprint,
 amsmath,amssymb,
 aps,
]{revtex4-2}

\usepackage{graphicx}
\usepackage{epstopdf}
\usepackage{dcolumn}
\usepackage{bm}
\usepackage{bbm}
\usepackage{xcolor}

\begin{document}

\preprint{APS/123-QED}

\title{Anomalous Floquet Topological Disclination States}

\author{Haoye Qin, Zhe Zhang, Qiaolu Chen, and Romain Fleury}
\affiliation{%
Laboratory of Wave Engineering, School of Electrical Engineering, EPFL, 1015 Lausanne, Switzerland}%

\date{\today}

\begin{abstract}
Recently, non-reciprocal two-dimensional unitary scattering networks have gained considerable interest due to the possibility of obtaining robust edge wave propagation in the anomalous Floquet phase.
Conversely, zero-dimensional topological states in such networks have been left uncharted. Here, we demonstrate the existence of Floquet disclination states in non-reciprocal scattering networks. The disclination states, characterized by spectral charges, nucleate in the anomalous phase from a resonant rotation-symmetric phase matching condition, and survive until the bandgaps close. Once coupled to the radiation continuum feeding the anomalous chiral edge state, they can induce intriguing topological disclination bound states in the continuum (BICs), associated with an extreme confinement and lifetime. Altogether, anomalous Floquet disclination states and topological disclination BIC broaden the applications of disclination states to microwave, acoustic or optical scattering networks, with new possibilities in chiral topological lasing, robust energy squeezing from topological bound states, and switchable lasing and anti-lasing behavior induced via unidirectional topological coupling.
\end{abstract}

\maketitle

Two-dimensional (2D) topological insulators are one of the simplest and practically relevant forms of topological systems, due to their compatibility with modern planar technologies and fabrication methods in electronics, photonics and phononics. They exhibit chiral or helical edge transport that originates from bulk-boundary correspondence, a property guaranteed by a quantized number known as bulk topological invariant \cite{qi_topological_2011,hasan_colloquium_2010}. Topological edge states have been realized in various platforms ranging from quantum to classical waves, including photonic and acoustic lattices \cite{olekhno_topological_2020,lu_topological_2014,he_acoustic_2016,xue_acoustic_2019}. In systems with broken time-reversal symmetry, Chern edge states have been observed and exploited for unidirectional wave routing and robust transport of energy \cite{haldane_model_1988,chen_tunable_2020,wang_observation_2009,ding_experimental_2019}. However, they are not the only possible 2D chiral edge states. In unitary (or Floquet) systems, for which the spectrum lives in a compact space, an extra anomalous topological phase exists, with zero Chern numbers for all bands, and an equal number of chiral edge states in each gap. While the topology of unitary systems is now well understood \cite{rudner_anomalous_2013,pasek_network_2014,delplace_phase_2017}, photonic anomalous phases have only been recently implemented in reciprocal \cite{rechtsman_photonic_2013,afzal_realization_2020} and even non-reciprocal unitary scattering systems \cite{zhang_superior_2021,zhang_anomalous_2023}. Surprisingly, anomalous edge transport was found to be much more resilient than Chern edge transport to various types of distributed perturbations, including arbitrarily large quasi-energy fluctuations, scattering matrix disorder and strong amorphism \cite{zhang_superior_2021,zhang_anomalous_2023}.

These years have also witnessed discoveries of higher-order topological states in topological insulators, like 0D corner or disclination states in 2D insulators \cite{xue_acoustic_2019,li_higher-order_2020,ni_observation_2019,liu_bulkdisclination_2021,chen_observation_2022,deng_observation_2022,peterson_trapped_2021,benalcazar_bound_2020,cerjan_observation_2020,hu_nonlinear_2021}, as well as hinge and corner states in 3D insulators \cite{schindler_higher-order_2018,benalcazar_quantized_2017}. Similar to topological edge-mode lasers \cite{bahari_nonreciprocal_2017,ota_active_2020,st-jean_lasing_2017,harari_topological_2018,bandres_topological_2018, xia_nonlinear_2021}, these 0D higher-order states lead to robust field trapping with stronger localization than 1D edge modes, enabling low-threshold topological lasing at corners \cite{kim_multipolar_2020,zhong_theory_2021}. Albeit disclination states have been extensively studied in topological crystalline insulators protected by time-reversal symmetry (TRS), the investigation of topological disclination states in TRS-breaking systems, including unitary scattering networks supporting the anomalous and Chern phases, still remain largely unexplored.

Here, we explore the existence of topological anomalous 0D states localized at disclinations in nonreciprocal scattering networks, composed of interconnected unitary circulators. Starting from a hexagonal lattice, a simple cut-and-glue procedure leads to 0D disclination states that nucleate in every bandgap, along with the 1D edge states. The occurrence of the disclination states at the phase-rotation symmetry points can be understood from a round-trip resonance condition and described by disclination charges. Remarkably, the disclination states do not disappear until the gaps close, for example at the Chern/anomalous phase transition point. To probe the disclination state, one can couple it with the edge state continuum, and create topological disclination bound states in the continuum (BICs), characterized by an infinite scattering condition number. The extremely high field enhancement, as well as the possibility to couple multiple disclination BIC via chiral edge modes reveal intriguing possibilities, including switchable amplified lasing and anti-lasing between two BIC sites.

\begin{figure*}[htbp!]
\includegraphics[width=1\textwidth]{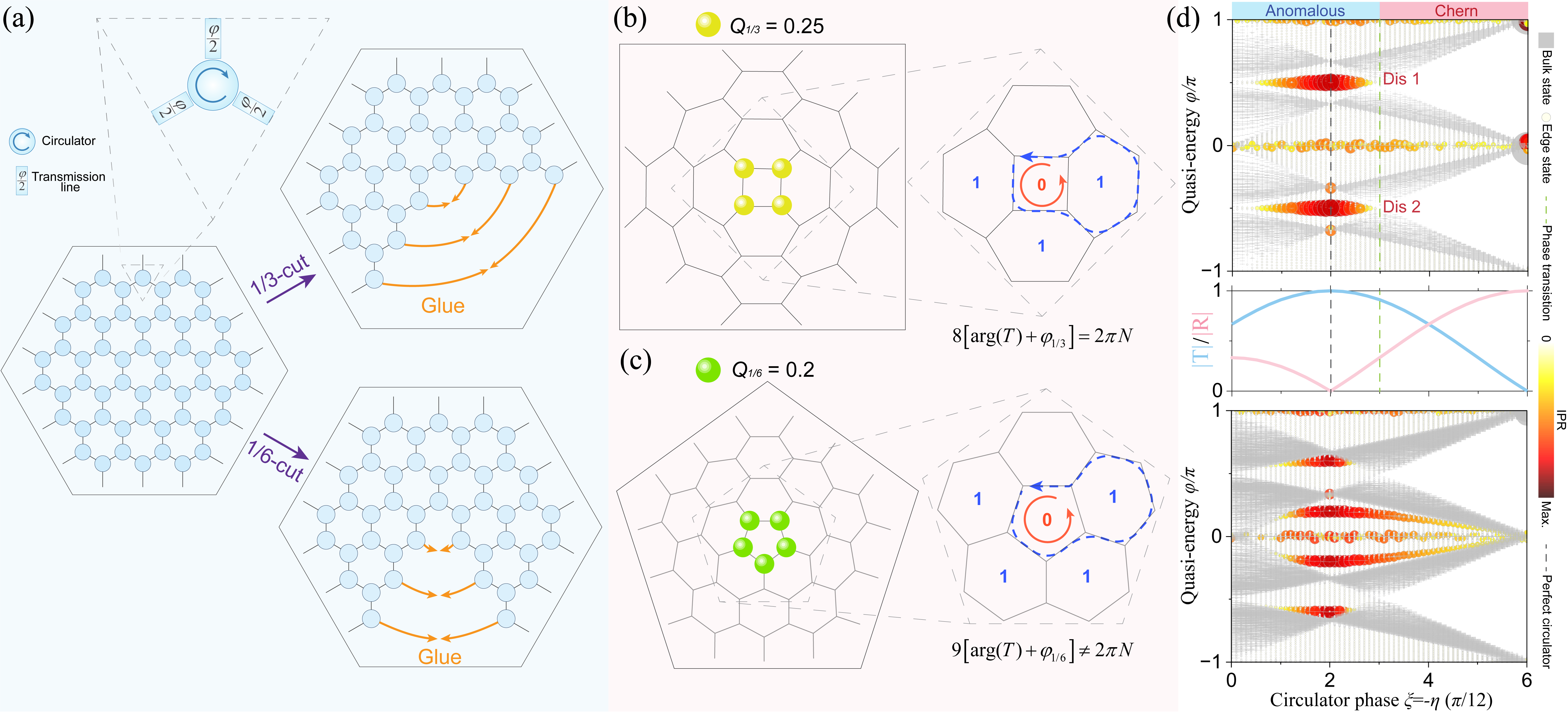}
\caption{\label{fig:1} \textbf{Anomalous disclination states in non-reciprocal scattering networks.} (a) Initial hexagonal network of circulators connected through transmission lines (TLs) with a phase delay $\varphi$, and cut-and-glue method for generating disclinations. (b, c) Disclination networks of 1/3-cut (b) and 1/6-cut (c) with distinct spectral charges (colored dots). The zoomed-in view of the center shows the principal disclination loop ``0" and surrounding higher order loops ``1". (d) Evolution of the quasi-energy band structure for the networks in (b) (top) and (c) (bottom) versus the circulator parameters $R,T$, plotted in the middle, which are controlled by a single circulator phase parameter. The color and size of each scatterer indicate the mode's inverse participation ratio (IPR), from grey (bulk states), light yellow (edge states) to red (disclination states).}
\end{figure*}

In Fig. 1(a), we consider a hexagonal non-reciprocal scattering network where each scattering node is a counter-clockwise circulator and each link is a monomode transmission line (TL) imparting a phase delay $\varphi$. The bulk modes supported by such scattering networks can be found by solving a unitary Bloch eigenproblem, where the eigenphase $\varphi$ belongs to a circle, therefore constraining the band structure to a compact space. Insulating phases in this lattice can then be classified as being either trivial, anomalous or of Chern type \cite{pasek_network_2014,delplace_phase_2017,zhang_superior_2021}. The Chern phase, as usual, has nontrivial Chern numbers that must add up to zero for all bands, thus exhibiting both trivial and topological bandgaps. In contrast, the anomalous phase with vanishing Chern numbers for all bands, hosts topological edge mode for every gap, dictated by a homotopy gap invariant $W$ \cite{delplace_phase_2017}. Beyond this theoretical distinction, one can physically distinguish the anomalous and Chern phases from the fact that anomalous edge transport is resilient to arbitrarily large quasi-energy and structural disorder, unlike the Chern phase \cite{zhang_superior_2021,zhang_anomalous_2023}. For $C_3$-symmetric circulators, this anomalous resilience is guaranteed when the magnitude of the circulator reflection coefficient $R$ is lower than $1/3$ \cite{zhang_superior_2021}.
The unitary scattering matrix of such circulators is $S_{c}=[R,T,D;D,R,T;T,D,R]$, where $T$ and $D$ are clockwise and counter clockwise transmissions, respectively. The parametrization of $S_{c}$ involves two independent phase parameters, $\eta$ and $\xi$, which control reflection $R$ and nonreciprocity $T/D$ \cite{zhang_superior_2021}. In the rest of this work, we set $\xi=-\eta$, which is sufficient to go from the anomalous phase ($0<\eta<\pi/4$) to the Chern phase ($\pi/4<\eta<\pi/2$) \cite{noauthor_see_nodate}. We now create disclination defects in such a system. Following a cut-and-glue method, a 1/3 (or 1/6) portion of the hexagon is firstly removed, before connecting the remaining TLs, keeping constant all phase delays in the
network. With perfect circulators ($R=0$, $\lvert T \rvert=1$, $\xi=-\eta=\pi/6$), chiral waves propagating along the disclination loop  (loop ``0" in Figs. 1(b) and 1(c)) are completely isolated from the surrounding loops, and the existence condition for a disclination state is the one of constructive self-interference after a round trip. Concretely, the quasi-energy $\varphi_{1/k}$ required for a $1/k$-cut disclination state is
\begin{equation}
\begin{aligned}
    n_k[arg(T)+\varphi_{1/k}]=2\pi N
\end{aligned}
\end{equation}
where $n_k=6(1-1/k)$ is the number of sides of the loop ``0", $T=-1$ is the transmission coefficient of perfect circulators and $N$ is an integer. This leads to $\varphi_{1/3}=\pm 0.5\pi$ and $\varphi_{1/6}=\pm 0.2\pi, \pm 0.6\pi$. We calculated the disclination spectral charge \cite{liu_bulkdisclination_2021} for each site to be $Q_{1/3}=0.25$ in Fig. 1(b) and $Q_{1/6}=0.2$ in Fig. 1(c). Summation of the charge over the four or five disclination sites always yields 1. Incidentally, the condition $R=0$ is known as a phase-rotation symmetry point \cite{delplace_phase_2017}, which occurs in the anomalous phase. Therefore, such disclination states are a consequence of the anomalous topology of the system. To evidence the presence of anomalous disclination states and investigate their evolution when moving away from the perfect circulator condition, we represent in Fig. 1(d) the band structure of the system as a function of the circulator phase $\eta=-\xi$. The color and size of each dot is proportional to the inverse participation ratio (IPR) of the mode. Bulk modes, with low IPR, are in grey. As expected, when $R=0$ ($\xi=-\eta=\pi/6$, black dashed line), one indeed observes the presence of disclination states at the expected values of $\varphi_{1/3}=\pm 0.5\pi$ (top) and $\varphi_{1/6}=\pm 0.2\pi, \pm 0.6\pi$ (bottom).  
When deviating from perfect circulators, we observe that the disclination 
states do not escape the bandgaps that host them: they only disappear at the bandgap closing condition $R=1/3$ ($\xi=-\eta=\pi/4$, blue dashed line), i.e. at the transition between anomalous and Chern phases, confirming their topological origin. In fact, for non-perfect circulators, the isolated disclination loop will be dispersed to generate higher order loops ``1", ``2", and so on, based on cycle basis in graph theory. Concentrating on loop ``1" but without loss of generality to other higher order loops, a surprising contrast occurs between the 1/3-cut and 1/6-cut disclination state: the former still satisfy the constructive self-interference condition at the same value of $\varphi$, which is not the case of the latter. This explains the fact that the 1/3-cut disclination states remains at a stable value of $\varphi$ away from the perfect circulator condition. Contrarily, the position of the 1/6-cut disclination state shifts, while remaining inside the bandgap. In both cases, the presence of the disclination state in a given gap coincides with a non-zero value of the homotopy gap invariant $W$, and the disclination state can be pictured as a small edge state looping around the center of the sample.

\begin{figure}[htbp]
\includegraphics[width=0.48\textwidth]{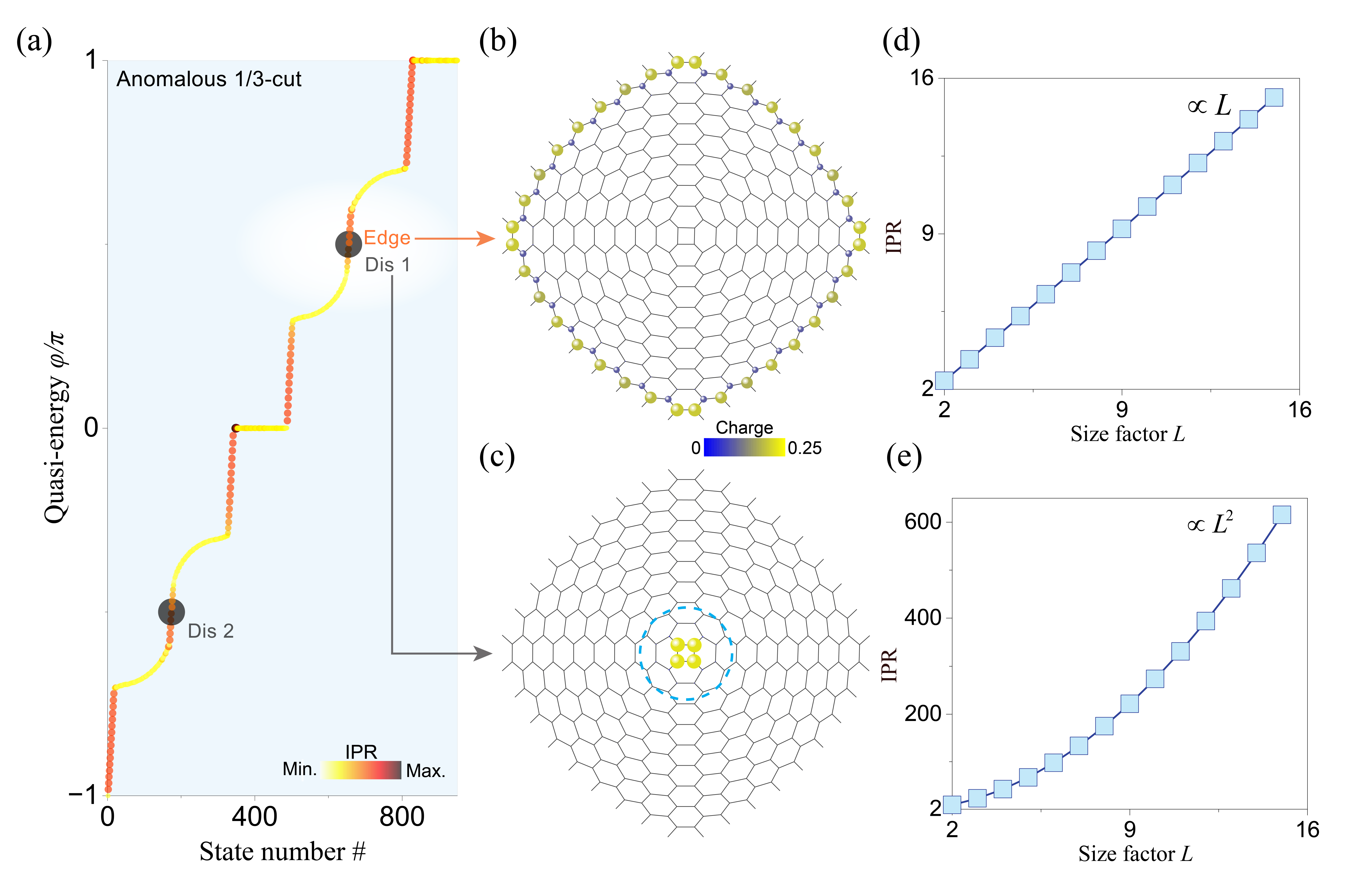}
\caption{\label{fig:2} \textbf{Different scaling properties of the inverse participation ratios of edge and disclination states.} (a) Disclination states with high IPRs can be obtained in an anomalous phase subject to a 1/3-cut. (b,c) Focusing on the anomalous case, we plot the spectral charge distribution for (b) edge (magnified by 10 times) and (c) disclination states ($\xi=-\eta=2.5\pi/12$). Both are chiral states propagating along either the outer or the inner tetragon. (d) IPR of the 1D edge state versus system size $L$. (e) Same as (d) for the disclination state.}
\end{figure}

\begin{figure}[htbp]
\includegraphics[width=0.47\textwidth]{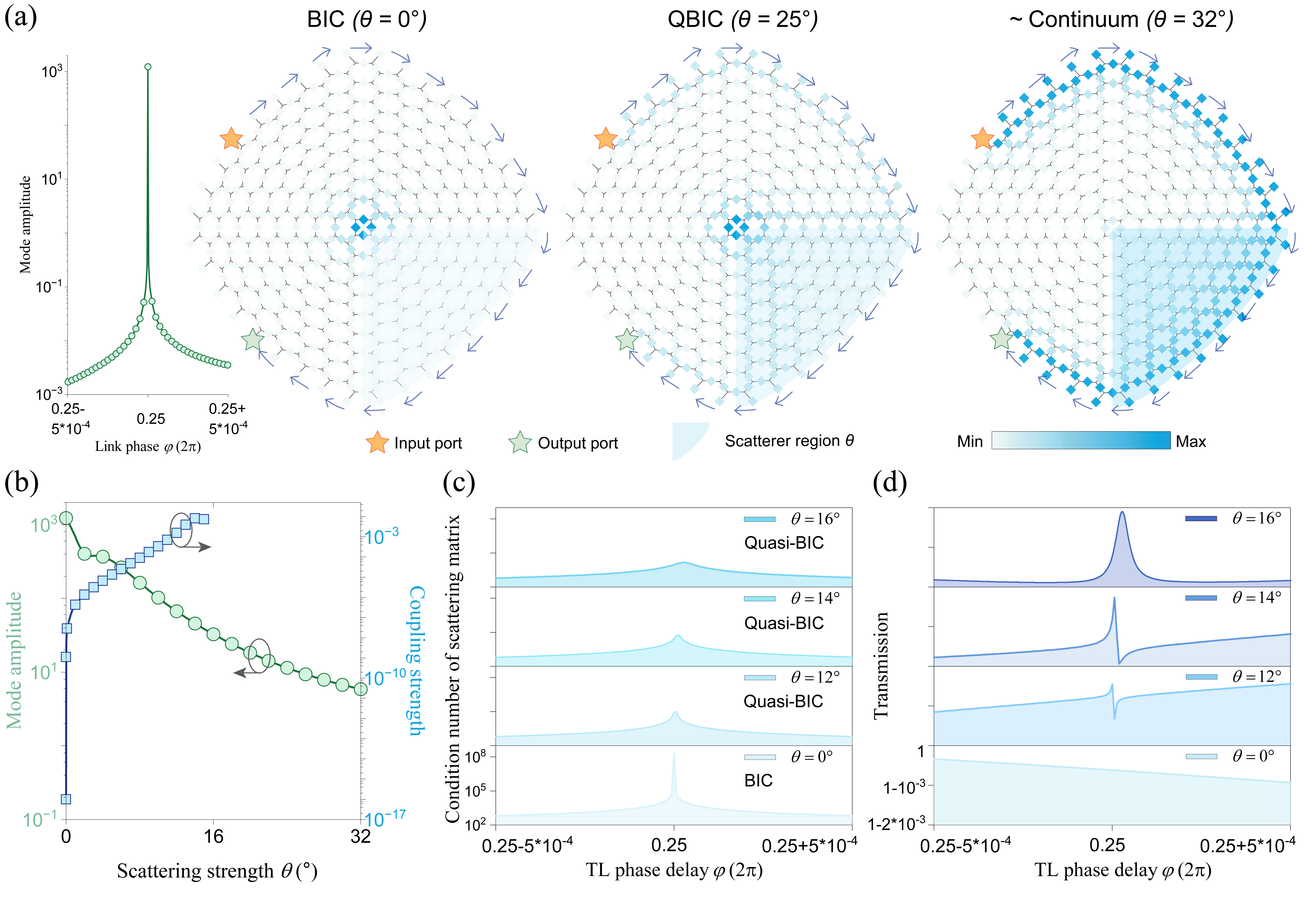}
\caption{\label{fig:3} \textbf{Topological disclination bound states in continuum.} (a) Layout and field for a 1/3-cut disclination system excited from edge ports. Scatterers are added to transmission lines in a quarter region, with tunable scattering strength $\theta$. Topological disclination BIC induces giantly amplified field intensity at the disclination site (inset). Increasing $\theta$ leads to stronger coupling with the continuum, undermining the BIC to a quasi-BIC. (b) Field amplitude and edge-disclination coupling strength of the disclination BIC versus $\theta$. Larger $\theta$ imposes stronger coupling $\langle\psi_{edge}|\psi_{dis}\rangle$ between the disclination and edge states, turning the disclination BIC into a quasi-BIC. (c) Condition number of the matrix $C_{net}-S_{net}$ versus phase for various scattering strength $\theta$. (d) Transmission from port 1 to port 2, confirming the vanishing linewidth associated with the BIC.}
\end{figure}

Fig. 2(a) elaborates on the differences between the 0D, 1D and bulk eigenstates of the disclinations samples. The figure focuses on a 1/3-cut in the anomalous phase. The color scale represents the IPR of the eigenstates, with darker color representing larger IPR. Disclination states have the largest IPRs, and are magnified as grey dots. One finds edge states everywhere (in red) except at the location of the bulk bands (in yellow), as expected for the anomalous phase. In the Chern phase, two trivial bandgaps are present, with no edge or disclination modes \cite{noauthor_see_nodate}. We plot the spectral charge distribution for an edge state and a disclination state in Figs. 2(b) and 2(c). Both are topological bound states with chiral energy flow. Since now the circulators are imperfect, the disclination charges leak outward, however an integration within the orange circle in Fig. 2(c) still yields the integer 1. Although the disclination state is intimately tied to the presence of an edge state, it is markedly different in terms of localization properties. Fig. 2(d) shows the evolution of IPR for the edge state when increasing the structure's size, which is proportional to the side length $L$. Conversely, for the disclination state, the IPR scales with $L^2$ (Fig. 2(e)), consistent with its zero-dimensional nature \cite{dominguez-castro_aubryandre_2019}. 
 
\begin{figure}[htbp]
\includegraphics[width=0.48\textwidth]{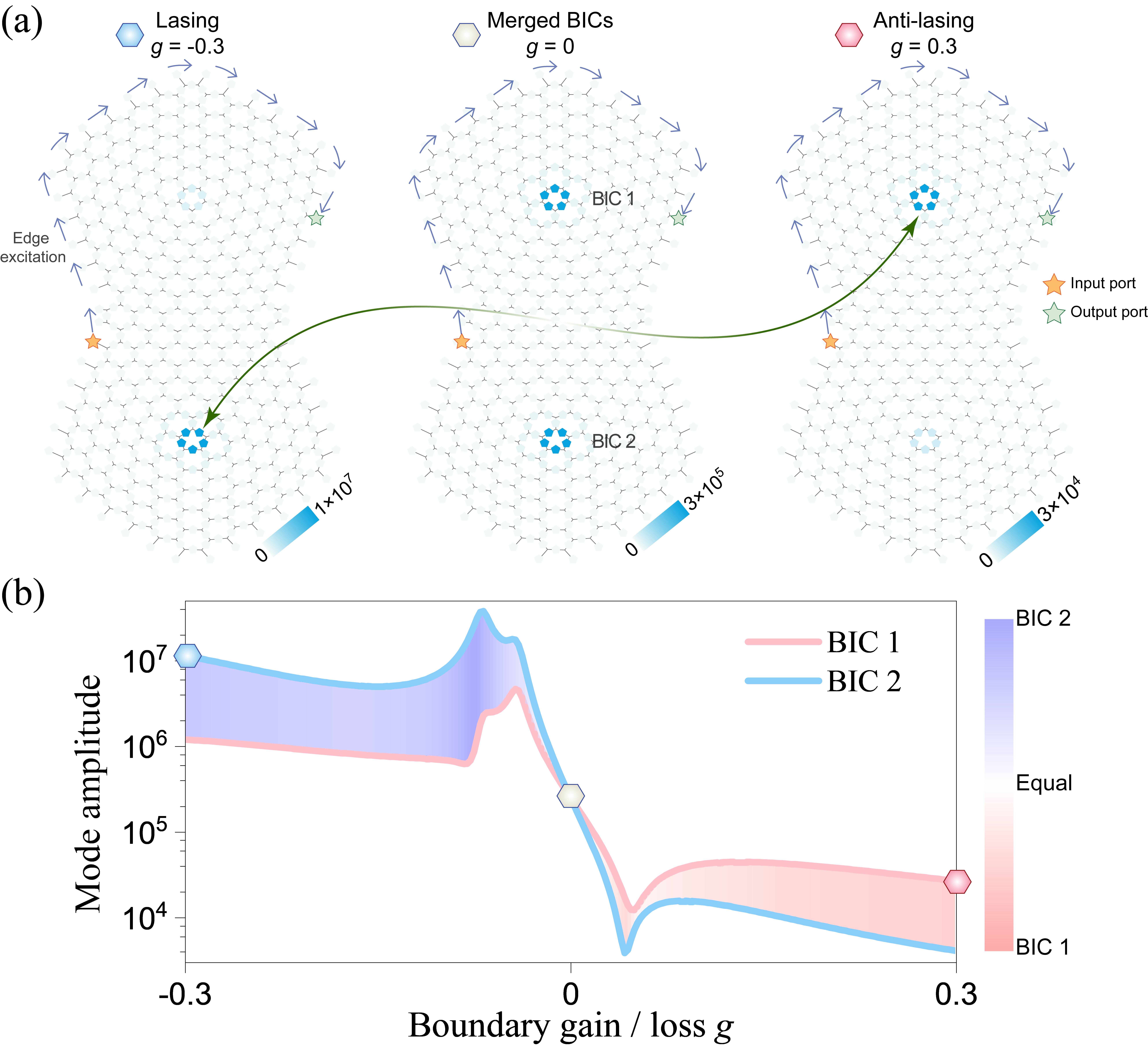}
\caption{\label{fig:4} \textbf{Lasing-anti-lasing switch by topological coupling of disclination BICs.} (a) Modulating gain ($g<0$) and loss ($g>0$) at the network boundary results in a switch between lasing and anti-lasing occurring when both BICs merge. In the Hermitian case ($g=0$), the two BICs have equal intensity. (b) Evolution of the field amplitude of the two BICs when tuning $g$. The markers correspond to the three cases in (a).}
\end{figure}

The fact that anomalous disclination states coexist within a broader continuum of edge states available from external ports can be exploited to create a unique form of robust BICs in which both the bound state and the connection to the continuum are of topological nature. By exciting the edge state from an input probe placed at the boundary (Fig. 3(a)), we can evanescently couple to the disclination state, generating a disclination BIC. The scattering matrix at the probes $S_{probe}$ is related to the internal degrees of freedom of the scattering network, and can be written as \cite{zhang_anomalous_2023}
\begin{equation}
    S_{probe}=[S_{ext}+S_{out}(C_{net}-S_{net})^{-1}S_{in}],
\end{equation}
where $S_{ext}$ is a unitary matrix describing the direct path scattering between the probes, $C_{net}$ is a unitary connectivity matrix describing the network, $S_{net}$ is a unitary matrix describing the scattering inside the network, and $S_{in/out}$ are non-unitary matrices describing the coupling between the ports and the network. This equation is the scattering network equivalent of the Mahaux-Weidenm\"uller formula usually used to express the scattering matrix of probes in terms of a Hamiltonian description of inner dynamics \cite{mahaux_and_h_a_weidenmuller_shell-model_nodate,zirnstein_bulk-boundary_2021}. When the matrix $C_{net}-S_{net}$ is not invertible, at least one inner site becomes fully disconnected and independent of the signal incident at the ports, which is the definition of a scattering network BIC. This BIC is associated to an infinite condition number of scattering matrix, and cannot couple to the continuum of topological edge states. 
Fig. 3(a) demonstrates the topological disclination BIC with remarkably enhanced field localized around the $1/3$-cut disclination, when the transmission line phase is tuned to be $\varphi_{1/3}=0.5\pi$. A sharp peak of mode amplitude occurs under this condition when sweeping the TL phase delay (inset). The field is extremely localized with a field enhancement factor of 1250 despite lying within the continuum channel, avoiding radiation. The topological disclination BIC can be checked by introducing the coupling with the continuum of edge states. Such coupling is induced by 2-port scatterers at the middle of the phase links in the shaded area shown in Fig. 3(a). These scatterers are described by 2-by-2 unitary scattering matrix with reflection coefficient $R= \sin\theta$, where $\theta$ defines a scattering strength. Tuning upward the value of $\theta$ from $0^\circ$ to $32^\circ$, the BIC (left, Fig. 3(a)) transitions to a quasi-BIC (center, Fig. 3(a)), and finally merges into the continuum in the strong coupling case (right, Fig. 3(a)). This transition is confirmed by increased coupling from nearly zero by several orders of magnitude (Blue line, Fig. 3(b)) and decreased mode amplitude (Green line, Fig. 3(b)) In addition, the condition number of $C_{net}-S_{net}$ in Fig. 3(c), initially infinite at $\theta=0$, transitions to smaller values. The peak of condition number corresponds to the peak of mode amplitude, which confirms that BICs in scattering network indeed arise when the matrix $C_{net}-S_{net}$ is not invertible. Sharp Fano-like signatures are obtained for quasi-BICs in the transmission spectrum measured at the output port (Fig. 3(d)), while tuning the scattering strength to zero also takes the spectral linewidth to zero, as expected for a BIC. Remarkably, the existence of the BIC is topologically protected, as verified by imparting disorder on $\varphi$ everywhere in the network \cite{noauthor_see_nodate}.
An extended application of this anomalous disclination BIC is shown in Fig. 4. By connecting two 1/6-cut disclination networks, a $y$-mirror symmetric octagon is formed with two disclination sites (top and bottom disclination). By introducing coupled BICs, we reveal isolated and merged phases differing from the BIC's individual disclination charges \cite{noauthor_see_nodate}. Near the perfect circulator condition ($\xi=-\eta=\pi/6$), the two BICs cannot couple via the central bulk and are therefore isolated, with the top disclination having a total charge of 1, and the bottom one a 0 charge, or vice versa. When deviating from this condition, a transition occurs as the two degenerate states start coupling, and they both end up with a disclination charge of 0.5, forming a merged BIC. We now start from this merged phase represented in Fig. 4(a), in which BICs have equal intensity, and add gain $g<0$ or loss $g>0$ on the boundary of the system. We see that the merged condition $g=0$ is the transition between lasing with gain and anti-lasing with loss. This is confirmed by the evolution of amplitude of two BICs in Fig. 4(b).  Differently, the isolated phase will enable BIC-enhanced symmetric and asymmetric lasing \cite{noauthor_see_nodate}. The advantage of this approach is that the amplitude of the disclination BIC can be robustly amplified in both the gain and loss regions, while bringing tunability and switchability in applications including robust lasing and anti-lasing at topological anomalous disclination BICs.

In conclusion, we explored the physics of anomalous Floquet topological disclination states in nonreciprocal scattering networks. By different cut-and-glue proportions, two disclination states with different stability in quasi-energy were revealed, both with markedly distinct inverse participation ratio scaling laws when compared to 1D topological edge states. When probing the anomalous disclination state from the edge state, we found that the topological zero-dimensional disclination state could form the basis for a robust bound state embedded in the topological edge state continuum, resulting in disclination BICs with direct topological protection against disorders. Using the largely enhanced field at disclination BICs, we demonstrated coupled disclination BICs that lead to intriguing phenomena like lasing switching between two disclination sites. Our study provides a new route for the exploitation of robust topological disclination states based on anomalous topology, and may contribute to novel possibilities to leverage topological BICs in non-Hermitian applications, in lasing and topological physics. 
  
\begin{acknowledgments}
This work was supported by the Swiss State Secretariat for Education, Research and Innovation (SERI) under contract number MB22.00028. Z. Zhang and R. Fleury acknowledge insightful discussions with Pierre Delplace about BICs in unitary scattering networks.
\end{acknowledgments}


\bibliography{referencesT}

\end{document}